\documentclass[fleqn,usenatbib,useAMS]{mnras}
\pdfoutput=1

\usepackage[nodayofweek]{datetime} 

\newdateformat{daymonthyeardate}{\THEDAY\;\monthname[\THEMONTH], \THEYEAR}
\usepackage{mathtools}
\usepackage{dcolumn}
\usepackage{bm}

\usepackage[utf8x]{inputenc}
\usepackage[T1]{fontenc}
\usepackage{enumitem}

\usepackage{amsmath}
\usepackage{amssymb}
\usepackage{graphicx}
\usepackage[colorinlistoftodos,color=lime,textwidth=21mm]{todonotes}
\usepackage[T1]{fontenc}
\usepackage{ae,aecompl}
\usepackage{newtxtext,newtxmath}
\usepackage{slashed}

\newcommand\WFIRST{{\slshape Roman}}

\title[LISA Galactic Binaries]{LISA Galactic Binaries in the {\slshape Roman} Galactic Bulge Time-Domain Survey}
\author[M. C. Digman, C. M. Hirata]{Matthew C. Digman$^{1,2,3,4}$\thanks{Contact e-mail: \href{mailto:matthew.digman@montana.edu}{matthew.digman@montana.edu}}, Christopher M. Hirata$^{1,2,5}$
\\
$^{1}$Center for Cosmology and AstroParticle Physics (CCAPP), Ohio State University, Columbus, OH 43210, USA\\
$^{2}$Department of Physics, Ohio State University, Columbus, OH 43210, USA\\
$^{3}$eXtreme Gravity Institute, Montana State University, Bozeman, Montana 59717, USA\\
$^{4}$Department of Physics, Montana State University, Bozeman, Montana 59717, USA\\
$^{5}$Department of Astronomy, Ohio State University, Columbus, OH 43210, USA}

\date{\daymonthyeardate\today}

\begin{document}
\label{firstpage}
\pagerange{\pageref{firstpage}--\pageref{lastpage}}
\maketitle

\begin{abstract}
Short-period Galactic white dwarf binaries detectable by LISA are the only guaranteed persistent sources for multi-messenger gravitational-wave astronomy.  Large-scale surveys in the 2020s present an opportunity to conduct preparatory science campaigns to maximize the science yield from future multi-messenger targets. The {\slshape Nancy Grace Roman Space Telescope} Galactic Bulge Time Domain Survey will (in its Reference Survey design) image seven fields in the Galactic Bulge approximately 40,000 times each. Although the Reference Survey cadence is optimized for detecting exoplanets via microlensing, it is also capable of detecting short-period white dwarf binaries. In this paper, we present forecasts for the number of detached short-period binaries the \WFIRST\ Galactic Bulge Time Domain Survey will discover and the implications for the design of electromagnetic surveys. Although population models are highly uncertain, we find a high probability that the baseline survey will detect of order $\sim5$ detached white dwarf binaries. The Reference Survey would also have a $\gtrsim20\%$ chance of detecting several known benchmark white dwarf binaries at the distance of the Galactic Bulge. 
\end{abstract}


\section{Introduction}

Short-period Galactic white dwarf binaries are the only guaranteed persistent targets for multi-messenger astrophysics with near-future gravitational-wave observatories. Such binaries are expected to be among the most numerous targets for the Laser Interferometer Space Antenna (LISA), a space-based mHz gravitational wave observatory scheduled to launch in the mid-2030s \citep{2017arXiv170200786A}. Because most such binaries will evolve minimally over the next 10-20 years, electromagnetic observations performed before LISA launches will help establish baselines for the evolution and behavior of such systems and enhance LISA's multi-messenger science yield \citep{Cornish:2017vip}. 

Multi-messenger observations of white dwarf binaries provide unique opportunities to study the astrophysics, formation, and tidal evolution of such systems \citep{2011CQGra..28i4019M,Carvalho:2022pst,2021arXiv211200145J,Yu:2020bic,2020ApJ...893....2L,2022arXiv220615390B,2021MNRAS.500L..52W}. Combining gravitational-wave and electromagnetic data for well-characterized systems can also be used to test for deviations from general relativity \citep{Littenberg:2018xxx,Xie:2022wkx,Wang:2022yxb} and augment LISA's calibration \citep{2018PhRvD..98d3008L}. Galactic white dwarf binaries are expected to produce a limiting foreground of gravitational-wave noise at mHz frequencies \citep{Timpano:2005gm,Edlund:2005ye,Nissanke:2012eh,Karnesis:2021tsh} which varies over time as the LISA antenna pattern rotates relative to the Galactic Center \citep{Digman:2022jmp}. Improved population modelling based on electromagnetic studies of these binaries can help to map the Galaxy \citep{Cornish:2001hg,Cornish:2002bh,Korol:2020lpq,Korol:2020hay,Roebber:2020hso,2022MNRAS.tmp.3426G,2022PhRvL.128d1101S,2020ApJ...901....4B}, and help separate the Galactic background form other potentially overlapping stochastic noise backgrounds including instrumental \citep{Adams:2010vc,2020PhRvD.102h4062E}, Galactic, extragalactic, and cosmological \citep{Adams:2013qma,2022PhRvD.105b3510B,2020CQGra..37s5020P} stochastic backgrounds  \citep{Bonetti:2020jku,2020MNRAS.491.4690M,Boileau:2021sni}. Because LISA data analysis inherently requires a global fit to the entire LISA data \citep{Cornish:2005qw,Crowder:2006eu,2009PhRvD..80f3007L,Littenberg:2020bxy}, improved characterization of the Galactic background benefits all LISA's gravitational-wave target searches including stellar origin black-hole binaries \citep{2022arXiv221204600D} and supermassive black-hole binaries, further augmenting the multi-messenger science yield. 
\pagebreak

Many current and near-future electromagnetic surveys will provide archival data of benefit to multi-messenger science with LISA, including but not limited to \WFIRST\  \citep{2019arXiv190205569A}, Rubin \citep{2012arXiv1211.0310L}, {\slshape Euclid} \citep{2011arXiv1110.3193L}, DES \citep{2005astro.ph.10346T}, DESI \citep{2013arXiv1308.0847L}, GAIA \citep{2008AN....329..875J}, ZTF \citep{2019PASP..131a8003M}, and TESS \citep{2014SPIE.9143E..20R}. In this work, we focus in particular on the potential of the planned Galactic Bulge Time Domain Survey (GBTDS) aboard the {\slshape Nancy Grace Roman Space Telescope} to detect a population of white dwarf binaries in the Galactic Bulge that are unlikely to be detected by any other near-future electromagnetic survey.  

The structure of this paper is as follows. In Sec.~\ref{sec:surveys}, we describe both the \WFIRST\ Galactic Bulge Time Domain Survey and LISA gravitational-wave observing mission. In Sec.~\ref{sec:methods}, we describe our methodology for using binary population synthesis models and simple instrument models to estimate the population of white dwarf binaries which can be detected by both LISA and \WFIRST. In Sec.~\ref{sec:known_systems}, we consider the mutual detectability of several systems that are already known to exist elsewhere in the galaxy. In Sec.~\ref{sec:results}, we present the results of our estimates for the population of mutually detectable binaries. In Sec.~\ref{sec:discussion}, we discuss the results and their limitations. In Sec.~\ref{sec:conclusions} we conclude and look towards future work. 

\section{The surveys}\label{sec:surveys}

\subsection{\WFIRST\ Galactic Bulge Time Domain Survey}

The \WFIRST\ mission, planned for launch in 2026, will carry a 2.4 m telescope to the Sun-Earth L2 Lagrange point. Its primary instrument, the wide-field instrument, will have 18 near-infrared detector arrays, each consisting of $4088\times 4088$ pixels at a scale of 0.11 arcsec per pixel, for a total field of view of 0.28 deg$^2$ \citep{2019arXiv190205569A}. Its filter wheel will carry 8 imaging filters and 2 dispersers for slitless spectroscopy. Because the solar arrays are fixed on the side of the observatory, the telescope can point anywhere in the range of 54--126$^\circ$ from the Sun.

The \WFIRST\ Galactic Bulge Time Domain Survey \citep{2019ApJS..241....3P} plans to make repeated observations of several fields in the Galactic bulge. The primary science objective of the GBTDS is to use microlensing to statistically characterize the population of extrasolar planets near or outside the Einstein radius of the lens stars by observing perturbations in the light curve. This includes statistical characterization of free-floating planets \citep{2020AJ....160..123J}. The fields will be observed in $\sim 72$ day long ``seasons'' during periods when the Galactic Bulge's orientation is favorable relative to the sun as seen from L2. In the reference observing scenario, these observations will be carried out mainly in the Wide filter (0.93--2.00 $\mu$m), which provides a higher photon count rate and, consequently, signal-to-noise than the standard-width imaging filters. The observatory will cycle through 7 neighboring fields, with an exposure time of $52\;$s per field, and complete a full cycle in $\sim900\;$s.\footnote{\citet{2019ApJS..241....3P} used an exposure time of 47 s; this has been increased to 52 s as of the August 2019 review of the \WFIRST\ Design Reference Mission.} We expect that $\sim 6$ such seasons will be possible in the 5-year \WFIRST\ primary mission, and more may become possible in an extended mission (there is propellant for $\ge 10$ years, and the infrared focal plane is passively cooled, so its lifetime will not be limited by cryogen). During bulge observations, the most important source of noise in most \WFIRST\ pixels will be Poisson noise from bulge stars (there are no empty pixels), followed by the zodiacal light and then instrument backgrounds and noise.

\WFIRST\ will take advantage of recent improvements in communications infrastructure and provide a data downlink rate of $\ge 11$ terabits per day, allowing it to downlink every pixel. This downlink rate makes \WFIRST\ imaging useful for both blind searches for variable sources and searches where the precise source position is not known in advance. The near-infrared detectors can be read non-destructively, so up to 6 sub-exposures out of each $52\;s$ exposure can be downlinked.\footnote{See the Science Requirements Document:
\url{https://asd.gsfc.nasa.gov/romancaa/docs2/RST-SYS-REQ-0020C\_SRD.docx}}

\subsection{LISA}

LISA is a constellation of three identical spacecraft orbiting the sun in a cart-wheeling triangle with 2.5 million kilometer arms \citep{2017arXiv170200786A}. Drag-free attitude control systems will hold the gold-platinum test masses at the end of each arm in near perfect geodesic motion \citep{Armano:2016bkm}. Laser links between the test masses will allow precise measurement of changes in the relative lengths of the arms. The technique of time-delay interferometery \citep{Tinto:2004wu} allows the individual phasemeter readouts from the laser links to be interfered in post-processing to construct polarization-sensitive observations of millihertz gravitational waves. 

LISA's nominal primary mission is four years, with a possible extension up to ten years. Because LISA sources can be observed continuously for years, far longer than is possible with ground-based gravitational-wave detectors, LISA will obtain comparatively good source localizations \citep{Cutler:1997ta}, opening opportunities to search for faint electromagnetic counterparts for a wide variety of sources. Detecting a source both electromagnetically and gravitationally will break numerous parameter degeneracies, in particular improving measurements of the distance to the source.  

The primary scientific observable for LISA is the strain in three time-delay interferometry channels as a function of time. While the exact distribution of sources between different classes depends on highly uncertain population modelling \citep{2022arXiv220501507S}, it is expected that LISA will have tens of thousands of sources observable at any given time, presenting a unique data processing challenge. The bulk of these sources will likely be Galactic white dwarf binaries \citep{Cornish:2017vip,Lamberts:2019nyk}, which will emit gravitational waves at a nearly constant source frequency for the duration of the LISA mission. 

Over a dozen already-known Galactic white dwarf binaries are guaranteed sources at LISA's design sensitivity \citep{Stroeer:2006rx,2018MNRAS.480..302K}. These are referred to as ``verification binaries.'' By the time LISA launches, several already-planned time-domain electromagnetic surveys will likely have discovered many more. 

\section{Methods}\label{sec:methods}

\subsection{Binary population synthesis}\label{ssec:bin_pop}
To generate the populations of binaries in the \WFIRST\ GBTDS, we use a modified version of COSMIC\footnote{\url{https://cosmic-popsynth.github.io/}} \citep{Breivik:2019lmt}, an updated version of BSE \citep{2002MNRAS.329..897H}. We generate a population of binaries with fixed metallicity $Z=0.018$ and a delta-function burst of star formation 8 Gyr ago for stellar ages.  Primary masses are drawn from a Kroupa IMF \citep{2001MNRAS.322..231K} with $m_1<100\;m_\odot$. Binaries with $m_1<0.95\;m_\odot$ are counted towards the total mass in the GBTDS field, but not evolved, because lighter primaries  never form compact objects with the physics encoded in COSMIC. We choose secondary masses uniformly from $0<m_2<m_1$, although we do not evolve systems with $m_2<0.5\;m_\odot$ because they also do not form compact binaries. We select initial eccentricities from a thermal eccentricity distribution, $f_e=2e$  \citep{1975MNRAS.173..729H}. We set the initial semi-major axis $a$ to be flat in log space from $0.5r_L<a<10^6r_\odot$, where $r_L$ is the Roche limit of the system. We use a modified-Mestel cooling law \citep{1952MNRAS.112..583M,Hurley:2003ku} for white dwarfs, and leave all other parameters to the COSMIC defaults. 

Higher multiple systems, such as hierarchical triple or quadruple systems, are beyond the scope of this work. It is possible that the existence of multi-star systems could produce a significant enhancement to the number of LISA-detectable short-period binaries, although blended starlight from brighter main-sequence companions may make identifying such systems in electromagnetic searches more difficult in some cases. 

After evolving the systems to the present, we discard all systems which are no longer binaries due to supernovae, mergers, or disruption events. We then discard all systems with final orbital period $P_f>10^5\;$s, which are unlikely to be LISA sources, recording only their final stellar mass. 

Based on the output of a run of TRILEGAL v1.6 \citep{2005A&A...436..895G, 2009A&A...498...95V, 2012ASSP...26..165G}, we expect a total mass of binary systems of $\sim1.5\times10^8\;m_\odot$ to be in the \WFIRST\  GBTDS fields after evolving the system. We calibrate the number of systems evolved with COSMIC to approximately match the expected mass in the field; drawing binaries from the initial mass distribution until $2\times 10^7$ pass our initial mass cuts gives approximately the correct total final mass of binaries.  

We then assign a new temperature to each binary which survives our output cuts for each version of our simplistic tidal heating model described in Sec.~\ref{ssec:tidal_heat}. For the radius of the system, we use the output core radius from COSMIC. Although the temperature and radius chosen this way are not generated from a self-consistent prescription, using the core radius at the lower temperature predicted by COSMIC should give a conservative estimate of the total luminosity of a system. Residual fusion and past accretion episodes are likely sources of additional heat which could further increase electromagnetic detectability, but such mechanisms are beyond the scope of this work. Once the temperature and radius are assigned, we calculate the atmospheric spectra as described in Sec.~\ref{sec:wd_atmospheres}. 

\subsection{Tidal heating}\label{ssec:tidal_heat}

We adopt the simple tidal heating prescription described in \citet{Burdge:2019hgl}. The contribution to the temperature due to tidal interactions alone, assuming tidal heat is instantaneously re-radiated, is given:
\begin{equation}\label{T_tides}
T_\text{tide}=\left(\frac{3\pi\kappa m}{2\sigma P_\text{orb}P_\text{crit} t_\text{gw}}\right)^{1/4},
\end{equation}
where $m$ is the mass of the white dwarf, $t_\text{gw}=3/2P_\text{orb}/|\dot{P}_\text{orb}|$ is the timescale for the binary to decay due to gravitational waves, $\kappa=I/(m r^2)$ is the coefficient of the moment of inertia, and $P_\text{crit}$ is some characteristic critical period which parametrizes the reduction in tidal heating as the spin period synchronizes to the orbital period. For simplicity, we adopt a fixed $P_\text{crit}=3600\;$s for our default model (see, e.g., Figure~4 of \citealt{2013MNRAS.430..274F}, also adopted by \citealt{Burdge:2019hgl}).  We use $\kappa\approx0.20306$, appropriate for an $n=1.5$ polytrope. 

Once we have the tidal contribution to the temperature, we add the temperature from the cooling age as predicted by COSMIC, $T_\text{eff}=(T_\text{cool}^4+T_\text{tide}^4)^{1/4}$. In practice $T_\text{tide}$ is always larger for the systems of interest here, so including the residual energy from cooling of the white dwarf binary makes little difference in $T_\text{eff}$. A recent history of accretion could make a more significant contribution to residual heat; the best-fit parameters for the recently detected J1539 \citep{Burdge:2019hgl} suggest it is a detached white dwarf binary, although it is much hotter than the cooling age or tidal heating alone could explain, suggesting a recent or ongoing accretion episode. Actively accretion is beyond the scope of this work, although such systems likely account for a large fraction of the systems that could realistically be detected by the \WFIRST\ GBTDS. 

Additionally, it is possible for low-mass He white dwarfs to maintain an outer hydrogen burning shell for much longer than a Hubble time, and therefore be much hotter than their cooling age would predict \citep{Althaus:2005jt,8176304,2016A&A...595A..45C}. Especially for younger white dwarfs, flashes of fusion can also heat and inflate the envelope, substantially enhancing the luminosity. Both effects are beyond the scope of this work.

\subsection{White dwarf atmospheres}\label{sec:wd_atmospheres}

For the white dwarf atmospheres, we interpolate a grid of precomputed spectra at various values of $T_{\text{eff}}$ and $\log(g)$ predicted by \citet{2010MmSAI..81..921K}\footnote{Precomputed spectra were downloaded from \url{http://svo2.cab.inta-csic.es/theory/newov/index.php?model=koester2}} for pure hydrogen atmospheres. We use a simple linear interpolation between the four nearest points to the requested $T_{\text{Eff}}$ and $\log(g)$. In the \WFIRST\ F146 filter, the spectra are effectively featureless such that interpolating this way produces an adequate approximation and properly accounting for the potential systems with hydrogen depleted atmospheres would have little to no impact on our results. For the handful of systems with parameters outside the precomputed grid, we default to a blackbody spectrum. In practice, our code does not predict any systems with a luminous primary with parameters outside the precomputed grid will be detectable for the Reference \WFIRST\ GBTDS survey. From the atmospheres, we obtain expected brightness for both systems, which are then used to generate light curves as described in Sec.~\ref{ssec:lc_gen}.

\subsection{Light curve generation}\label{ssec:lc_gen}

Light curves for eclipsing white dwarf binaries are generated using ELLC \citep{2016A&A...591A.111M}. ELLC models the white dwarfs as triaxial ellipsoids with a Roche potential \citep{1979ApJ...234.1054W}. We treat neutron stars and black holes as spherical. We adopt a linear limb darkening law with fixed coefficients $\text{ldc}_1=\text{ldc}_2=0.5$. For simplicity we do not include gravity darkening, Doppler boosting, or heating by companions, although all three effects could be significant. 

For light curve generation, we also do not include a period derivative, as including it would not affect the detection efficiency. 

We also set the eccentricity $e=0$ for light curve generation. In the absence of multi-star interactions, appreciably nonzero ($e>0.1$) eccentricities in LISA's band are expected to occur only for systems which contain a neutron star or black hole which has received a large natal kick from a supernova. Systems with COSMIC predicted eccentricities $e>0.1$ contribute $<0.1\%$ to our predicted number of systems detectable by LISA+\WFIRST, such that accounting for eccentricity would make no difference to our reported results. 

After generating the light curve with ELLC, we apply a dust extinction correction to the light curve as described in Sec.~\ref{ssec:dust_extinction}. Finally, we draw the actual observed electron counts for a given exposure as described in Sec.~\ref{ssec:wfirst_exposure}.

\subsection{Dust extinction}\label{ssec:dust_extinction}

For the dust extinction as a function of wavelength, we use the $R_V=3.1$ extinction curve from \citet{Weingartner:2000cc}\footnote{Downloaded from \url{https://www.astro.princeton.edu/~draine/dust/dustmix.html}}.

For the normalization of the extinction correction, we use the $A_H$ map \citep{2012A&A...543A..13G} used for planning the \WFIRST\ GBTDS \citep{2019ApJS..241....3P}, which uses $\lambda_H=1.64\;\mu\text{m}$ as the central wavelength for the H-band filter \citep{2012A&A...537A.107S}\footnote{The map is available at \url{https://github.com/mtpenny/wfirst-ml-figures/fields/GonzalezExtinction.txt}. Note the extinctions in the map are $A_K$ and must be rescaled using $A_H=1.559322A_K$.}. We linearly interpolate this extinction map when drawing points.

To draw random points in the GBTDS field, we obtain the chip centers from the Reference Survey.\footnote{See: \url{https://github.com/mtpenny/wfirst-ml-figures/fields/layout_7f_3.chips}} We then assign each realization of a binary to a random chip and select a uniformly random point in ($l$,$b$) on the chip to draw its $A_H$ from the dust map. At present, the $A_H$ drawn this way is uncorrelated with the background noise on an individual pixel drawn in Sec.~\ref{ssec:wfirst_exposure}, which may overestimate background noise for deeply extincted fields somewhat. 

Because we apply the dust extinction correction after generating the light curve, we do not allow the extinction to alter the brightness ratio of the systems. Within the range of white dwarfs of interest for this study, extinction is not a strong enough function of temperature for correctly perturbing the brightness ratio to affect our conclusions.

\subsection{\WFIRST\ Exposures}\label{ssec:wfirst_exposure}

For the \WFIRST\ exposures, we use the \WFIRST\ F146 filter\footnote{Effective areas for the F146 filter were obtained from \url{https://wfirst.gsfc.nasa.gov/science/201907/WFIRST_WIMWSM_throughput_data_190531.xlsm}}. From the effective area $A_\text{F146}(\nu)$, we fix the normalization of received light to electrons detected per second using
\begin{equation}
\dot{N}_\text{sig}(t)=f_\text{ap}10^{-0.4 m(t)}\int_0^\infty \frac{F^\text{ref}_\nu}{h\nu}A_\text{F146}(\nu)d\nu,
\end{equation}
where $F^\text{ref}_\nu=3.631\times10^{-20}\;erg/cm^2/s/Hz$, and $m(t)$ is the light curve in magnitudes as described in Sec.~\ref{ssec:lc_gen}. We consider signal photons hitting a $3\times3$ grid of adjacent pixels, assuming a constant aperture correction of $f_\text{ap}=0.815$. A more optimal choice of aperture is beyond the scope of this work.

To obtain the expected total count of electrons observed by a single \WFIRST\ exposure of length $dt_\text{exp}$, we compute a $dt_\text{exp}$ width rolling integral of $\dot{N}_\text{sig}(t)$ for a single orbit of the system and interpolate to obtain the expected electron counts at the mid-exposure time specified by the survey strategy described in Sec.~\ref{ssec:survey_strat}.

The expected number of signal electrons $N_\text{sig}(t,dt_\text{exp})$ for an exposure is then added to the expected background $N_\text{bg}(dt_\text{exp})$ for the exposure. For each system, we draw the background count rate $\dot{N}_\text{bg}$ from a random $3\times3$ grid of pixels of a single simulated exposure as in \citet{2013MNRAS.434....2P} (see also \citealt{2019ApJS..241....3P}). We then draw the observed electron counts from a Poisson distribution with $N_{\exp}(t,dt_\text{exp})=N_\text{sig}(t,dt_\text{exp})+N_\text{bg}(dt_\text{exp})$.

We then add a Gaussian random read noise of $\sigma_\text{count}=60$ electrons (i.e., 9 pixels with $\sigma_\text{pixel}=20$ electrons/readout/pixel adding incoherently) to the number of electrons actually observed $N_\text{obs}$. In most cases the read noise dominates the detectability limit for the sources of interest. To reduce the effective read noise, we assume the Fowler 2 readout scheme is applied, which averages two adjacent $2.88\;\text{s}$ detector readouts to reduce the effective read noise in the down-linked electron count rates by a factor of $\sqrt{2}$, at the expense of also reducing the effective length of each exposure by a single $2.88\;\text{s}$ sub-exposure. The process is then repeated for every exposure to obtain the full light curve observed by \WFIRST, which is then processed as described in Sec.~\ref{ssec:chisq}.

To account for saturation, we mask all pixels with  $\dot{N}_\text{bg}>20000\;$e/s, and reduce our quoted detection efficiencies by a factor of the fraction of pixels masked, $f_\text{mask}$.

\subsection{Survey Strategy}\label{ssec:survey_strat}

For the \WFIRST\ GBTDS, we consider a single 72 day survey with $n_\text{visit}=6912$ visits to each field, or visiting with a $P_\text{visit}=900\;$s cadence. For each visit, we take $n_{\rm exp}$ exposures of length $t_{\rm exp}$ (e.g., $n_{\rm exp}=1$, $t_{\rm exp}=52\;$s).

To reduce aliasing for orbital periods close to small integer fractions of $P_\text{visit}=900\;$s, we add a Gaussian random variation to the time between visits $\sigma_\text{visit}=6\;$s, which in the real survey could correspond to variation in the time required to acquire the guide stars upon arriving at a field.
(It is also possible that one would insert deliberate pseudo-random variations in the observing strategy, since this would reduce aliasing artifacts in any search for periodic signals in the bulge survey, e.g., in astroseismology; \citealt{2015JKAS...48...93G}.)

\subsection{Chi-squared calculation}\label{ssec:chisq}

For a given run, we first phase fold the simulated \WFIRST\ light curve into $n_{\text{bins}}=\text{Round}\left[P_\text{orb}/dt\right]$ phase bins with centers linearly spaced in phase on a grid $\phi'_\text{orb}\in[0,0.5]$, where we fold about the midpoint of the light curve $\phi_\text{orb}\equiv\text{mod}\left[t/P_\text{orb},1\right]$:
\begin{equation}
\phi'_\text{orb}=\begin{cases}
\phi_\text{orb} & \phi_\text{orb}\le0.5\\
1-\phi_\text{orb} & \phi_\text{orb}>0.5\\
\end{cases}.
\end{equation}

Folding about the midpoint of the light curve picks out sin-like variation, because we expect the variation from eclipses and ellipsoidal variation to be sin-like. Note that our folding strategy assumes the initial phase of the binary can be effectively fit.  Each sample is then placed in the phase bin whose center is closest to the phase at the midpoint of the exposure, and the $\chi^2$ is calculated:
\begin{equation}\label{chi2}
\chi^2=\sum_{i=0}^{n_{\text{bins}}}\frac{\left(\bar{N}_i-\bar{N}\right)^2}{\sigma_i^2},
\end{equation}
where $\bar{N}_i$ is the mean number of electrons in the $i$th phase bin, and $\bar{N}$ is the mean number of electrons over the entire light curve. The variance $\sigma_i^2$ can be estimated in  $n^\text{samp}_i$ exposures out of $n^{\text{exp}}$ total exposures with electron the electron count in each sequential exposure given $N_\alpha$:

\begin{align}
\sigma_i^2 &= \frac{\sigma_{lc}^2}{n^\text{samp}_i}\nonumber\\
&=\frac{1}{n^\text{samp}_i}\left[\frac{1}{n^{\text{exp}}}\sum_{\alpha=0}^{n_{\text{exp}}}\left(N_\alpha-\bar{N}\right)\right].
\end{align} 

Using this estimator of $\sigma_i^2$ corresponds mathematically to testing the null hypothesis that the variation in the light curve is stochastic. For a strongly signal dominated light curve, it will overestimate the variance compared to other possible estimators, such as the theoretically expected variance in a bin with $n_i^\text{samp}$ electrons. However, because using the empirical variance absorbs all experimental sources of variance in the count rate, it is more robust than using a theoretical variance.  

We then evaluate the significance according to a $\chi^2$ distribution, with the number of degrees of freedom $k=n'_\text{bins}-1$, where $n'_\text{bins}$ is the number of bins with $n_i^\text{samp}>0$. $n'_\text{bins}$ can be less than $n_\text{bins}$ due to aliasing, although aliasing can be mitigated by avoiding exactly periodic visits as described in Sec.~\ref{ssec:survey_strat}.

\subsection{Selecting binaries}\label{ssec:selecting_binaries}

We count a system as detectable by LISA when the four-year sky and inclination averaged signal to noise $S/N\ge7$, as defined by \citep{Cornish:2018dyw}. This condition is approximate, but should be sufficient to estimate the population of detectable binaries. We assume the frequency evolution over the course of the mission is small enough that it can be ignored for the purposes of calculating the signal to noise. 

We count a system as detected by \WFIRST\ when $p<10^{-10}$, based on the $\chi^2$ discussed in Sec.~\ref{ssec:chisq}. The small $p$ cutoff is necessary to avoid false positives due to the large number of pixels being searched. In order to be conservative, we calculate the detectability of a system given only a single 72-day Bulge viewing season; including the full campaign would increase the detection efficiency.

In a more theoretically optimal treatment, one could use the LISA sky localizations of high S/N LISA sources to reduce the trials factor for the \WFIRST\ search, allowing detection with a less conservative $p$ cutoff. For this purpose, we are most interested in sources which can be detected with reasonable high significance \emph{before} LISA launches, to facilitate using longer time baselines for computing period derivatives and electromagnetic characterization. However, LISA characterization will improve significantly even if only the position of a faint electromagnetic counterpart can be determined. Therefore \WFIRST\ archival data can facilitate ongoing improvement in the localizations of some LISA systems even if the \WFIRST\ GBTDS fails to detect them at sufficiently high significance on its own. 

\subsection{Trials factor}\label{ssec:trials_factor}
To compute the trials factor, we assume the search runs over linearly spaced frequency bins of width $\Delta f=1/(2 t_\text{season})\approx 8\times10^{-8}\;\text{Hz}$ from $f_\text{min}=10^{-4}\;\text{Hz}$ to $f_\text{max}=1/(2 t_\text{exp})\approx9.6\times 10^{-3}\;\text{Hz}$, resulting in $n_f=118,387$ possible frequency bins. Each frequency bin is divided into phase bins of width $\Delta\phi=\text{Round}\left[t_\text{exp}f\right]$, resulting in a total of $n_\text{bins}=1,090,156$ frequency+phase bins.  The seven GBTDS fields will contain $n_\text{pix}=2.016\times 10^9$ pixel locations to search.  Therefore the total number of trials will be $n_\text{trials}\approx2.2\times10^{15}$. Because we report single season detection efficiencies, candidate binaries in the first GBTDS season can be verified in the second season; therefore, in two seasons the expected number of false positives is given $\left<n_\text{fp,2}\right>\approx2.2\times10^{15}p_\text{cut,1}^2$, where $p_\text{cut,1}$ is the threshold for identifying a candidate detection in one season. If $\left<n_\text{fp,2}\right>=10^{-4}$ is the maximum tolerable number of false positives, then we require $p_\text{cut,1}\approx 2.1\times10^{-10}$.

\subsection{Calculating detection efficiency}\label{ssec:monte_carlo}
To calculate the detectability of a system in \WFIRST\ output by our
Each realization of a binary system is given a uniformly random initial phase, as well as a pixel background noise as described in Sec.~\ref{ssec:wfirst_exposure} and a dust extinction as described in Sec.~\ref{ssec:dust_extinction}. All binaries in the \WFIRST\ GBTDS fields get the same realization of the variation in exposure timing descrived in Sec.~\ref{ssec:survey_strat}. For each binary we calculate the $f_\text{detected}(i)$ at each point on a grid of $n_\text{incl}$ inclination bins linearly spaced in $\cos i\in\left[0,1\right]$. 

We then integrate to obtain the probability of detection:
\begin{equation}
f_\text{detected}=\int_0^1f_\text{detected}(i)\,\,{\rm d}\cos i.
\end{equation}

For each inclination bin, we do $n_\text{run}$ realizations of the systems to compute $f_\text{detected}(i)$. 

\section{Known systems}\label{sec:known_systems}
As of this writing, ZTF J153932.16 +502738 (J1539), SDSS J065133.338+284423.37, and ZTF J2243+5242 (J2243) are the three shortest period known detached white-dwarf binaries \citep{2011ApJ...737L..23B,2012ApJ...757L..21H,Burdge:2019hgl,Burdge:2020bul,Burdge:2020end,Brown:2020uvh}. All three will have LISA $S/N\gtrsim90$ at their observed spectroscopic distances of $d\simeq2.3\;\text{kpc}$, $d\simeq1.0\;\text{kpc}$, and $d\simeq2.1 \;\text{kpc}$ respectively \citep{Littenberg:2019mob}.  Because $S/N\propto1/d$, all three would still be easily detected as LISA sources ($S/N$>7) in the Galactic Bulge at $d\simeq8\;\text{kpc}$. J1539 and J0651 sources are believed to be He-CO binaries. For J0651, the He white dwarf is hotter, and consequently the primary by luminosity, and the system is detected primarily by the eclipses of the He white dwarf by the CO secondary. J2243 is most likely He-He binary. For J1539, the CO white dwarf primary is extremely hot ($T_\text{eff}\simeq48,900 K$) and is detected due to both very strong reflection/reprocessing of light by the otherwise unseen secondary, and total eclipses of the primary.

Even after artificially adding our tidal heating prescription in post-processing, the physics encoded by COSMIC cannot produce a short-period He-CO binary with a super-heated CO white dwarf primary such as J1539. At present, it is impossible to know for certain how representative J0651, J2243, and J1539 are of the general binary white dwarf population. Because Nature actually produced these systems, they represent useful benchmarks for the detectability with \WFIRST. It is of course possible that the locally observed white dwarf binary population could be quite different from the old stellar population in the Galactic Bulge. We do note however that the location of J1539, 1.8 kpc from the Galactic Plane \citep{Burdge:2019hgl}, is consistent with an older halo population (the other possibility being that J1539 is much younger and experienced a large kick due to mass ejection in a binary interaction; e.g., \citealt{2020ApJ...890...69L}). In this work, we consider the detectability of J0651, J2243, and J1539-like sources in more detail. 

For J1539, reflection presents an additional complication because the spectrum for a $T_\text{eff}\simeq48,900 K$ white dwarf is very blue. The measured CHIMERA g'-band reflective heating coefficient is $\text{heat}_2=3.851^{+0.159}_{-0.147}$ \citep{Burdge:2019hgl}. Because $\text{heat}_2>1$, most of the light is actually reprocessed UV light, rather than true reflected light, and the reprocessing is wavelength-dependent. Modelling the physics of such reprocessing is beyond the scope of this work. Therefore, to give a sense of the range of plausible \WFIRST\ F146 light curves for J1539, we present two scenarios: one with the best fit g'-band reflection coefficient, $\text{heat}_2=3.851$, and one with no reflection, $\text{heat}_2=0$. 

Additionally, we sample a grid of possible inclinations for both sources, evenly spaced on $\cos i\in[0,1]$. For each inclination we then generate 100,000 random realizations of the light curves with as described Section~\ref{sec:methods}. The other binary parameters are set to the appropriate values from \citet{2012ApJ...757L..21H} and \citet{Burdge:2019hgl}.

\section{Results}\label{sec:results}

\begin{figure}
\includegraphics[width=\columnwidth]{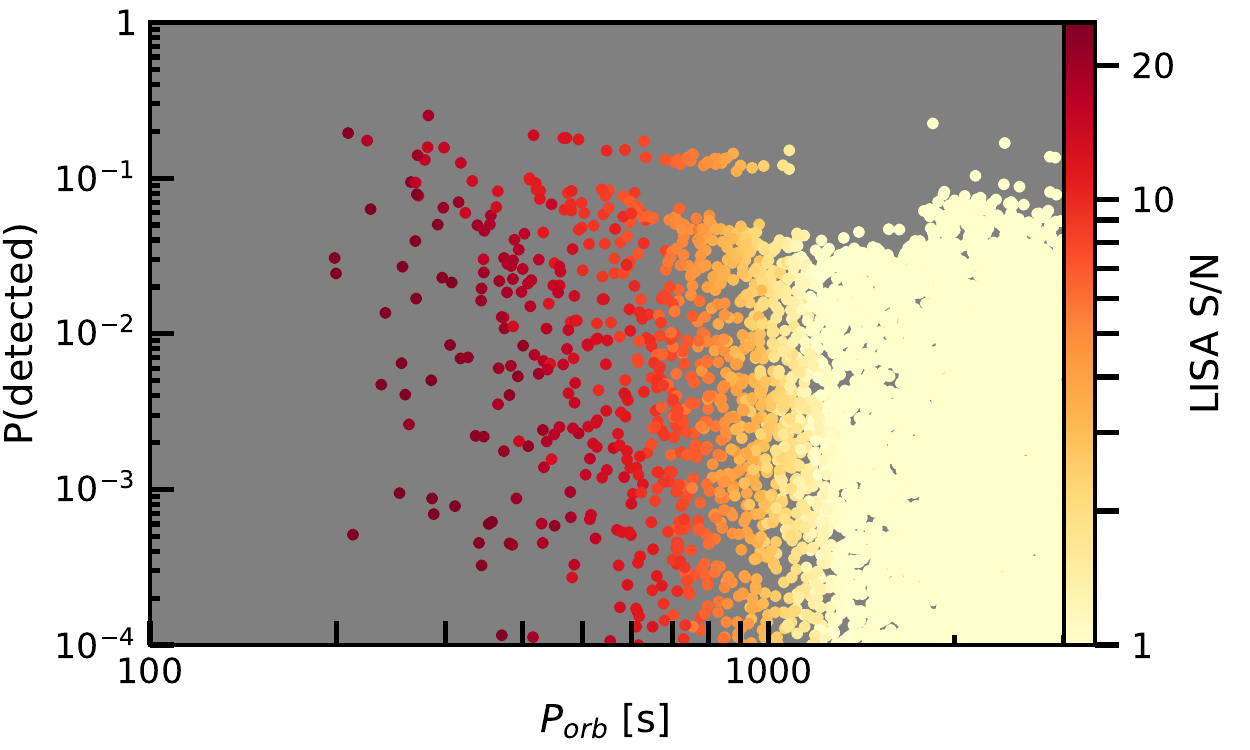}
\caption{\WFIRST\ detection probability versus period for our fiducial run with temperatures from cosmic, with estimated LISA S/N shown in color. The higher mass of the He-CO binaries gives them higher LISA S/N at fixed period, while the electromagnetic luminosity of the CO white dwarfs is generally smaller than He white dwarfs, resulting in the apparent inverse correlation between LISA S/N and \WFIRST\ detection efficiency at fixed period. The small population of binaries separated from the overall trend in the top middle is a population of systems with very young (age$<5\;\text{Myr}$), hot secondaries.}\label{fig:mutual_det_no_tide}
\end{figure}

\begin{figure}
\includegraphics[width=\columnwidth]{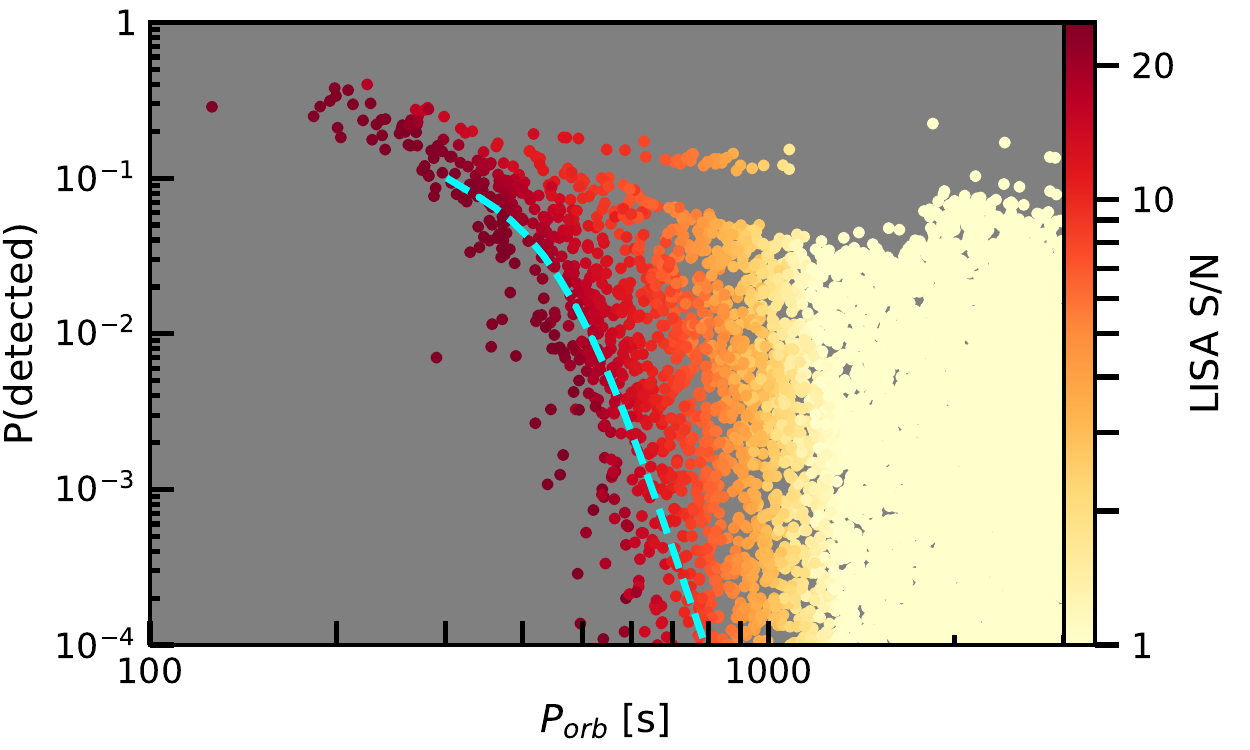}
\caption{\WFIRST\ detection probability versus period for our fiducial run with tides enhanced to match the temperatures of the J0651 system, with estimated LISA S/N shown in color. For a given period, approximately 95\% of He-He binaries fall above the dashed line. In this model, all He-He binaries with $P_f\lesssim680\;$s have a $>0.3\%$ chance of being detected in a single season of the nominal \WFIRST\ GBTDS. For He-CO binaries, all systems with $P_f\lesssim475\;$s have a $>0.3\%$ chance of being detectable by \WFIRST. }\label{fig:mutual_det_standard_tide}
\end{figure}

\begin{figure}
\includegraphics[width=\columnwidth]{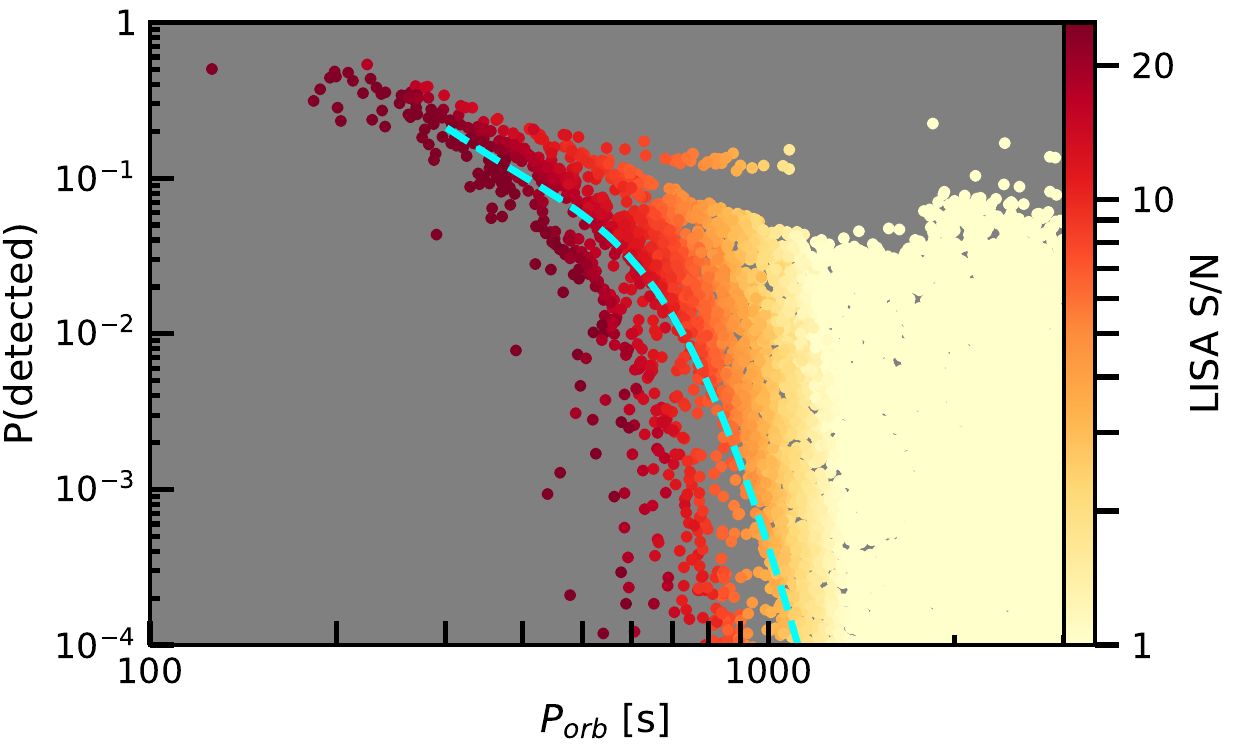}
\caption{\WFIRST\ detection probability versus period for our fiducial run with tides enhanced to match the temperatures of the J0651 system, with estimated LISA S/N shown in color. For a given period, approximately 95\% of He-He binaries fall above the dashed line. In this model, all He-He binaries with $P_f\lesssim950\;$s have a $>0.3\%$ chance of being detected in a single season of the Reference \WFIRST\ GBTDS. For He-CO binaries, all systems with $P_f\lesssim475\;s$ have a $>0.3\%$ chance of being detectable by \WFIRST.}\label{fig:mutual_det_enhanced_tide}
\end{figure}

In Figs.~\ref{fig:mutual_det_no_tide}, \ref{fig:mutual_det_standard_tide}, and \ref{fig:mutual_det_enhanced_tide}, we present the results from our fiducial run, summarized in Table~\ref{tab:expectations}. The most detectable single system generated by any run of our models is a $P_{orb}=224.4\;$s He-He binary with a $41,000\;K$ primary and a $39,000\;K$ secondary generated in by our model with artificially enhanced tidal heating, which would have $f_\text{detected}\simeq 57\%$ if it were in one of the \WFIRST\ GBTDS fields.  No individual system with an orbital period $P_{orb}>370\;$s has $f_\text{detected}>25\%$, but collectively the large number of systems with individually low chances of having an inclination favorable enough to be detectable are the dominant contributions to the expectation values quoted in Table~\ref{tab:expectations}, rather than the handful of systems with individually high $f_\text{detected}$.

\begin{center}
\begin{table}
\caption{Expectations values for number of binaries detected with $P_\text{orb}<3000\;$s computed based on 20 Monte Carlo realizations of the 7 GBTDS fields$^\text{\ref{foot:exp_uncertainty}}$. The quoted $1-\sigma$ errors are purely statistical uncertainties on the expectation values, and are far smaller than the inherent systematic uncertainties in the formation, evolution, and modelling of these systems. }\label{tab:expectations}
\begin{tabular}{ l l l l l }
\hline\hline
 Type & Tide Model & \WFIRST\ & LISA & \WFIRST+LISA \\ 
\hline
Total & No Tides       & $3.07\pm0.08$ & $50.3\pm1.6$ & $0.43\pm0.04$\\
 Total & Basic Tides    & $4.33\pm0.12$ & $50.3\pm1.6$ & $1.54\pm0.10$\\
 Total & Enhanced Tides & $6.23\pm0.18$ & $50.3\pm1.6$ & $2.89\pm0.16$\\
 He-He & No Tides       & $2.99\pm0.08$ & $26.2\pm1.1$ & $0.42\pm0.04$\\ 
 He-He & Basic Tides    & $3.91\pm0.11$ & $26.2\pm1.1$ & $1.20\pm0.09$\\
 He-He & Enhanced Tides & $5.50\pm0.16$ & $26.2\pm1.1$ & $2.24\pm0.14$\\
 He-CO & No Tides       & $0.05\pm0.01$ & $19.9\pm1.0$ & $0.01\pm0.00$\\
 He-CO & Basic Tides    & $0.38\pm0.05$ & $19.9\pm1.0$ & $0.33\pm0.05$\\
 He-CO & Enhanced Tides & $0.70\pm0.08$ & $19.9\pm1.0$ & $0.64\pm0.08$\\
\hline\hline
\end{tabular}
\end{table}
\end{center}

\footnotetext{\label{foot:exp_uncertainty}Expected total number of systems from 20 realizations of the Galactic Bulge Time Domain Survey is approximated using $\left<n\right>\simeq\sum_i f^i_\text{detected}/20$, with 1-sigma statistical error on the mean approximated using $\sigma_{n}\simeq\sqrt{\sum_i \left(f^i_\text{detected}\right)^2}/20$}

\subsection{Benchmark Systems}\label{ssec:benchmark}

\begin{figure}
\includegraphics[width=\columnwidth]{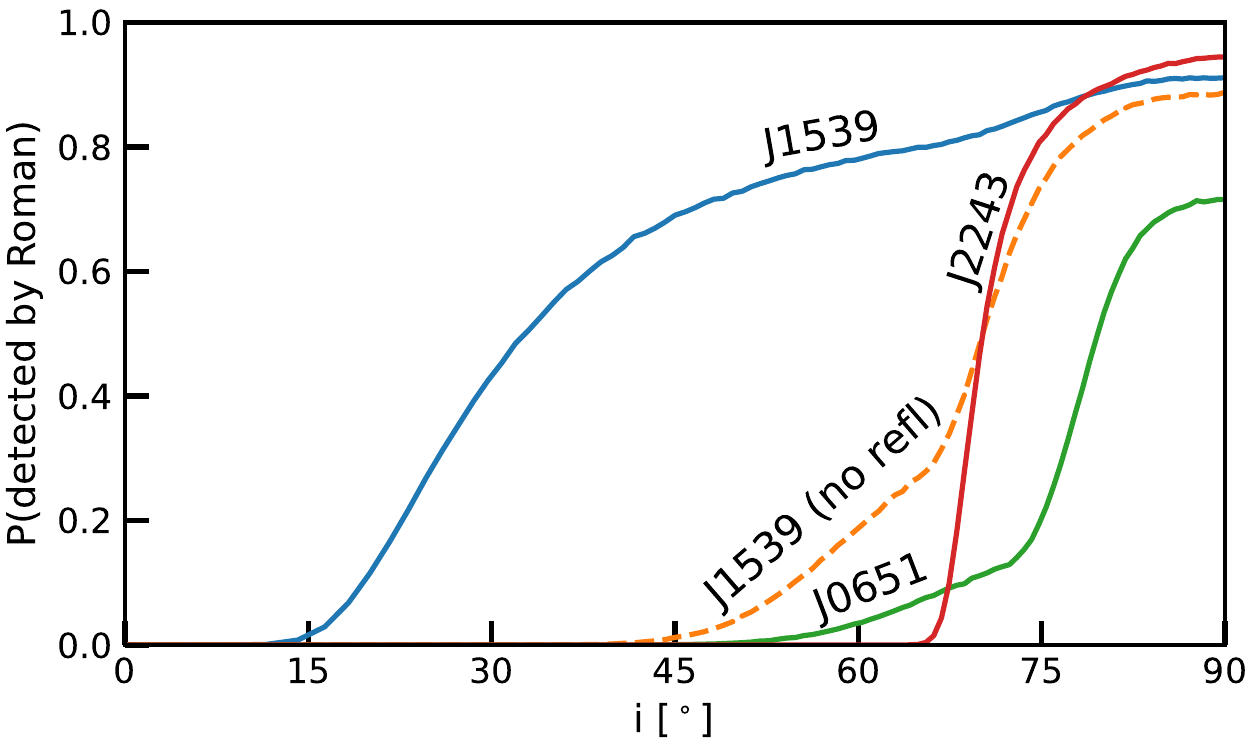}
\caption{\WFIRST\ detection probability versus inclination for our three benchmark systems at the distance of the Galactic Bulge. J1539 is shown both with and without the observed reflection effect, to give an idea of how much the reflection contributes to the detectability. For inclinations where it would be eclipsing, the candidate He-He white dwarf binary J2243 is the most detectable system. The He-CO binaries exhibit more ellipsoidal variations and therefore are detectable to lower inclinations, especially if reflection is also enabled.}\label{fig:incl_var}
\end{figure}

Our forecasts for the \WFIRST\ detectability of our three benchmark systems as a function of inclination are shown in Fig.~\ref{fig:incl_var}. All three systems have a substantial probability of being detectable if they were present at the distance of the Galactic Bulge in the \WFIRST\ GBTDS fields. The detectability is strongest at high inclinations. These systems are more detectable than the sources that contribute to the rate estimates presented in Table~\ref{tab:expectations}. If these benchmark sources are representative of typical sources in the Galactic Bulge, the prospects for joint detections could be much better than reflected in the rate estimates.

Overall, our model of the He-He binary J2243 has an inclination-averaged \WFIRST\ detection probability of $\sim31\%$ at the assumed distance of the Galactic Bulge of 8,000~$\text{pc}$, and the He-CO binary J0651 has an inclination averaged detection probability of $\sim17\%$. The J1539-like system has an inclination averaged $\sim70\%$ probability of detection including our reflection model, and $\sim34\%$  including only ellipsoidal variation and eclipses.

\section{Discussion}\label{sec:discussion}

In this work, we have only considered statistical false positives due to Poisson fluctuations in background light and detector readout noise. We have endeavored to  set an aggressive single-season $p$ value requirement of $p<2.1\times10^{-10}$ to guarantee that virtually no statistical false positives will remain as candidate detached white dwarf binaries after two GBTDS seasons. This requirement is likely very conservative and represents candidate binaries that should be quickly and straightforwardly identified and likely represent the best candidates for follow up with other instruments. A more optimized Bayesian analysis of a simulated dataset for the entire \WFIRST\ GBTDS would likely return a higher expected number of true positives while maintaining a low rate of statistical false positives, although such candidates may be more difficult to validate with other instruments. 

Validation with other instruments is an essential consideration because the bulk of high signal-to-noise false positives in the actual \WFIRST\ GBTDS survey will likely be astrophysical. There are several other types of astrophysical signals that could masquerade as binaries with a comparable period. Rotating stars with spots could exhibit light curve variations on similar periods, as could asteroseismology or pulsations. While such systems may be interesting in their own right, they are contaminants for the purposes of this study. Forecasting the number and period distribution of such systems identifiable with the \WFIRST\ GBTDS is beyond the scope of this paper. In most cases, LISA itself will be able to validate the identification of a periodic source as a binary, especially for periods $\lesssim900\;\text{s}$ where virtually all binaries of white dwarf mass objects will be detectable by LISA at $8\;\text{kpc}$. Other electromagnetic instruments, especially those with higher resolution or larger collecting areas, such as next-generation extremely large telescopes, could follow up candidate binaries to make a more definitive classification. Follow-ups can also provide additional valuable scientific information of value to multi-messenger studies, such as colors and temperatures. Sources that can be identified by both multiple electromagnetic instruments and LISA are appealing multi-messenger science targets.

\section{Conclusion}\label{sec:conclusions}

In this work, we have presented several forecasts of the number of short-period detached white dwarf binaries detectable by the {\slshape Nancy Grace Roman Space Telescope}'s planned Galactic Bulge Time Domain Survey based on populations of such binaries simulated using COSMIC. We find that it is probable that at least a handful of detached white dwarf binaries will be detected by the planned survey, and that most such binaries will also be detectable by LISA. The expected number of binaries identified is a strong function of the assumed temperatures of the binary components, which relies on physics such as tides, residual fusion, and accretion history. This physics is not included in the population synthesis code we have used. Additionally, we have excluded binaries undergoing mass transfer from our analysis, which will be significantly heated and, therefore, potentially represent a larger fraction of the total detectable systems. Mass transferring and detached systems are both separately interesting \citep{2021ApJ...908....1S}. Ideally, both types of systems can be detected so their evolutionary properties can be compared.    

In addition to uncertainties in the temperatures, uncertainties in population synthesis as a whole are very large \citep{Korol:2021pun}, and will not be substantially reduced until future generations of surveys begin and place observational constraints on formation channels. Because the Galactic Bulge is a different stellar population from the local population where the bulk of currently-known detached short-period white dwarf binaries have been detected, multi-messenger detections of sources there have the potential to improve our understanding of binary formation channels and stellar astrophysics. Indeed, several observed short-period binaries are not well-predicted by existing temperature models \citep{2020ApJ...890...69L}, and as shown in Sec.~\ref{ssec:benchmark}, the detectability of those systems would be quite high in the \WFIRST\ GBTDS. Consequently, it is not possible to make a precise forecast of the number of detached white dwarf binary systems that will be detectable by \WFIRST. We have attempted to capture some of this uncertainty by presenting three artificial models of tidal heating, which produce results representing a range of point estimates within the realm of plausibility. 

The planned \WFIRST\ GBTDS is not optimized to detect short-period white dwarf binaries. To detect short-period periodic sources, it would be more efficient to conduct a continuous series of exposures in each field, rather than spending a significant fraction of the total survey time slewing and settling. The data gaps associated with the survey cadence also introduce aliasing effects that cause the survey to lose efficiency at certain periods. Aliasing effects can also be mitigated by taking a continuous series of exposures on a single field. Such considerations could motivate using a small fraction of the GBTDS survey time allocation to conduct such a series of exposures, which might also be valuable for other scientific purposes within the survey. A survey campaign specifically optimized for identifying multi-messenger science targets in the Galactic Bulge could also be proposed and conducted in a possible \WFIRST\ extended mission after the primary mission has been completed. Observations conducted in the first few planned \WFIRST\ GBTDS seasons will be useful for constructing a better-calibrated forecast of the number of systems an optimized survey campaign would be capable of finding.

We did not conduct any detailed LISA instrument modelling in this work, including modelling the effect of the time-variability of the Galactic stochastic background, which can be a significant effect for sources close to the Galactic Center \citep{Digman:2022jmp}. A more detailed analysis of the impact of combining LISA and \WFIRST\ data could better illuminate the effect of mutual detection of sources on the LISA global fit.

Multi-messenger science is an exciting and rapidly expanding field. Planned survey instruments will be capable of identifying valuable multi-messenger targets. Early identification of some of those targets with \WFIRST\ will be a scientifically valuable tool to predict and optimize the multi-messenger science yield achievable in concert with LISA.

\section*{Acknowledgements}

We would like to thank Matthew Penny for useful discussion and providing many of the files related to the \WFIRST\ GBTDS, and Samaya Nissanke for useful discussions. MCD and CMH were supported by the Simons Foundation award 60052667, NASA award 15-WFIRST15-0008, and the US Department of Energy award DE-SC0019083. MCD was also supported by NASA LISA foundation Science Grant 80NSSC19K0320. The computations in this paper were run on the CCAPP condo of the Pitzer Cluster at the Ohio Supercomputer Center \citep{Pitzer2018}.

\section*{DATA AVAILABILITY}

The data underlying this article will be shared on reasonable request to the corresponding author.

\bibliographystyle{mnras}
\bibliography{lisa_wd}

\bsp	
\label{lastpage}
\end{document}